\documentclass[]{scrartcl}
\usepackage[utf8]{inputenc}
\usepackage{authblk}
\usepackage[numbers,sort&compress]{natbib}
\usepackage{hyperref}
  \hypersetup{colorlinks,linkcolor=[rgb]{0.2,0.2,0.75},citecolor=[rgb]{0.3,0.5,0.5},urlcolor=[rgb]{0.4,0.4,0.6},breaklinks}
\usepackage{doi}
\usepackage{amsmath}
\usepackage{graphicx}
\usepackage{fancyhdr}
\usepackage{xcolor}
\usepackage{soul}
\usepackage{bm}
\usepackage{siunitx} 
\usepackage[margin=10pt,font=small,labelfont=bf,labelsep=endash]{caption}
\usepackage{float}

\setkomafont{title}{\raggedright\fontsize{22}{26}\selectfont\sffamily\bfseries}

\makeatletter
\renewcommand{\maketitle}{%
  \begin{flushleft}
    {\usekomafont{title}\@title\par}
    \vspace{1.5em}
    {\@author\par}
    {\@date}
  \end{flushleft}
}
\makeatother

\title{JOREK simulations of the X-point radiator formation and its movement in ASDEX Upgrade}
\date{}

\author[1,2]{Y. C. Liang $^*$}
\author[1]{A. Cathey}
\author[1]{M. Hoelzl}
\author[3]{S. Q. Korving}
\author[1,2,3]{M. Szucs}
\author[1]{O. Pan}
\author[4]{D. Maris}
\author[1,2]{F. Antlitz}
\author[$\ddagger$]{the JOREK Team}
\author[$\mathsection$]{the ASDEX Upgrade Team}

\affil[1]{Max Planck Institute for Plasma Physics, Garching b. M. and Greifswald, Germany}
\affil[2]{TUM School of Natural Sciences, Physics Department, Garching, Germany}
\affil[3]{ITER Organization, St. Paul Lez Durance Cedex, France}
\affil[4]{DIFFER, Eindhoven, the Netherlands}
\affil[$^*$]{Email: \href{mailto:yu-chih.liang@ipp.mpg.de}{yu-chih.liang@ipp.mpg.de}}

\newcommand{\reprintinfo}[1]{}

\begin{document}

\maketitle

\begingroup
\renewcommand{\thefootnote}{\fnsymbol{footnote}}
\footnotetext[3]{\hypertarget{jorek}{}see author list of M. Hölzl \textit{et al.} 2024 \textit{Nucl. Fusion} \textbf{64} 112016}
\footnotetext[4]{\hypertarget{asdex}{}see author list of H. Zohm \textit{et al.} 2024 \textit{Nucl. Fusion} \textbf{64} 112001}
\endgroup

\section*{Abstract}

Future large-scale magnetic confinement fusion reactors require operational regimes that can avoid extreme heat fluxes onto the plasma-facing components. One promising regime is the X-point radiator (XPR), which relies on a highly radiative, cold and dense plasma volume forming above the X-point, and which can be accessed via impurity seeding. Experimentally, the height of the XPR can be controlled by adjusting the seeding rate and heating power. This contribution presents axisymmetric (2D) simulations of the XPR regime in ASDEX Upgrade using the nonlinear MHD code JOREK extended with a kinetic particle framework for the main species neutrals and nitrogen impurities. With the time-dependent simulations, the progression from attached divertors to a complete detachment with the XPR formation is shown, highlighting the effects of the neutrals and impurities separately. Amidst this progression, the formation and the loss of the high-field-side high-density are observed. After the XPR is well-formed at the height of 6.8\;cm, the fuelling and seeding rates are adjusted so that the XPR remains stationary. From the stationary case, the seeding rate is then changed to see how the XPR location reacts. By increasing and decreasing the seeding rate, the XPR responds by moving upwards and downwards, respectively. These simulations show JOREK’s capability of simulating time-varying XPR, which will provide a baseline for the transition to 3D simulations, so the MHD activities and their interaction with the XPR can be studied.
\section{Introduction}\label{intro}

For a reactor-scale tokamak, strict management of the power exhaust is required to maintain the heat flux densities to the material structures in the divertor region at low levels, in order to guarantee a sufficient component lifetime~\cite{Exhaust_Pitts_2019_NME}. On the one hand, the quasi-stationary heat loads must be controlled by dissipating the vast majority of large heat fluxes via radiation, which distributes them over a broader surface in comparison to the narrow divertor strike-lines that would be wetted otherwise~\cite{Exhaust_Loarte_2007_NF}. On the other hand, the enormous transient heat loads associated to bursts of 3D magneto-hydrodynamic (MHD) instabilities like periodic type-I edge localized modes (ELMs)~\cite{ELM_Zohm_1996_PPCF} are not tolerable in a fusion power plant. Therefore, it is crucial to achieve exhaust scenarios without type-I ELMs avoidance and with divertor detachment, while maintaining core plasma performance.~\cite{Exhaust_Viezzer_2023_NME}.

The \textit{X-point radiator} (XPR) operational regime promises a way of achieving divertor detachment via impurity seeding. The XPR is a highly radiative, cold, and dense plasma volume that can be stabilised above the active X-point. Its vertical position can be controlled via tuning the impurity seeding rate or the heating power~\cite{XPR_control_Bernert_2021_NF}. Similar to a MARFE\footnote{Multifaceted asymmetric radiation from the edge}~\cite{MARFE_Lipschultz_1984_NF}, the formation of an XPR is due to radiation condensation~\cite{MARFE_Drake_1987_POF}. A cold XPR core has the electron temperature reduced to around 1.5\;eV mainly due to the impurity radiation, and the local electron density increases to around 10$^{20}$\;m$^{-3}$ to maintain the parallel pressure balance along the magnetic field lines. At this temperature, the XPR can be maintained stable and stationary above the X-point. Only when the temperature is further reduced, the XPR loses plasma density via strong recombination and turns into an unstable MARFE by drifting towards the inner midplane (IMP)~\cite{XPR_model_Stroth_2022_NF, XPR_control_Sieglin_2025_FED}. This transition typically leads to a disruption. The distinction between the two phenomena is further described in Ref.~\cite{XPR_Bernert_2025_NME}. In ASDEX Upgrade (AUG), an XPR can be kept stable up to approximately 15\;cm above the X-point, and it can be maintained stationary by a real-time feedback control system, with the nitrogen seeding rate as the actuator. Additionally, by obtaining a high XPR position (typically 7\;cm in AUG), an ELM-suppressed regime can be reached~\cite{XPR_control_Bernert_2021_NF}, with the energy confinement time only marginally reduced.

While experimental and theoretical work has investigated the basic principles of XPR formation well and also provided insights regarding the suppression of ELMs, there is still a lack of understand on the underlying processes behind the XPR movement and the interaction between the XPR and the ELMs. The work at hand aims to provide insight into the XPR movement by simulating the dynamics of XPR formation, loss, and its transition into a MARFE. The simulations were conducted using a fluid-kinetic hybrid approach, using the MHD code JOREK~\cite{JOREK_Hoelzl_2024_NF} and treating the neutral particles and impurities with a full-f particle-in-cell method~\cite{JOREK_kinetic_Korving_2024_phd}. The focus of this work is to study the formation and the development of an XPR, as an axisymmetric phenomenon, in relation to the deuterium neutrals fuelling and the nitrogen seeding. In this article, axisymmetric (2D) simulations are presented, and these simulations will also provide a basis for future work, including 3D effects like the ELMs and experimental validation.

The rest of the article is structured as follows: Section~\ref{setup} introduces the JOREK code, the hybrid fluid-kinetic model, and the experimental AUG scenario forming the basis for our work. The different phases of the simulation setup are also described. Section~\ref{formation} addresses the dynamics related to the initial formation of an XPR. Section~\ref{ref} establishes a reference scenario after the XPR has become quasi-stationary. Section~\ref{marfe} and~\ref{retreat} then investigate the resulting dynamics if the impurity seeding rate is increased further or reduced, respectively. Section~\ref{collisions} addresses the additional aspects introduced by the neoclassical collisions of the impurity particles with the fluid background. Section~\ref{conclusion} summarizes the findings, provides conclusions from them and outlines future work that should be addressed.


\section{Experimental scenario and simulation setup}\label{setup}

To initiate the reference simulation presented in this article, the magnetic equilibrium and the plasma conditions of AUG discharge~\#38773 at 3.65\;s\footnote{Time traces of the discharge parameters can be found in Ref.~\cite{SOLPS_Pan_2023_NF}, figure 1.} were used. In this discharge, the plasma current was 0.8\;MA, and the toroidal magnetic field was $-$1.8\;T, with the ion grad-B drift pointing to the X-point. The total plasma heating power was 10\;MW, whereas the total radiative power corresponded to around 7.5 MW. The deuterium fuelling rate was kept constant at 2\;$\times$\;10$^{22}$\;e$^-$/s, whilst the nitrogen seeding rate was feedback-controlled to move the XPR to the preset position, and it was also around 2\;$\times$\;10$^{22}$\;e$^-$/s at the 3.65\;s time point. This time point also corresponded roughly to when the XPR formed.

For the simulations presented in this article, the non-linear extended MHD code JOREK~\cite{JOREK_Hoelzl_2024_NF, JOREK_Hoelzl_2021_NF, JOREK_Huysmans_2007_NF}, coupled with the fully kinetic extension for neutral and impurity particles described in Refs.~\cite{JOREK_kinetic_Korving_2023_POP, JOREK_kinetic_Korving_2024_POP, JOREK_kinetic_Korving_2024_phd, JOREK_kinetic_Vugt_2019_phd}, was used. Previously, with this kinetic particle framework, a simplified JOREK XPR simulation was achieved for the first time and presented in Ref.~\cite{JOREK_kinetic_Tongeren_2023_master}.

For the background deuterium plasma, the reduced MHD model was used, in which the toroidal field compression is neglected, and the model is single-temperature, including the $E$$\times$$B$, grad-B, and diamagnetic drifts. The kinetic framework models both the deuterium neutral~\cite{JOREK_kinetic_Korving_2023_POP} and the nitrogen impurity~\cite{JOREK_kinetic_Korving_2024_POP} particles as a Monte-Carlo particle tracker, which is two-way coupled to the background MHD fields, and a \textit{full-f} approach is followed. In this framework, particles are pushed based on the local MHD fields with a Boris integrator~\cite{JOREK_kinetic_Boris_1970, JOREK_kinetic_Delzanno_2013_JCP}. In the presented simulations, the two species of kinetic particles enter the simulation from the private flux region (PFR), since in AUG, the valves for gas puff fuelling and nitrogen seeding are also situated in the PFR~\cite{Detachment_Henderson_2023_NF}. To model the interactions between the kinetic particles and the background MHD fields, the atomic reactions included are ionisation, charge exchange, volumetric recombination, and the effective lines and continuum radiation~\cite{JOREK_kinetic_Korving_2023_POP}, which is calculated with the effective collisional radiative rates from the OpenADAS database~\cite{ADAS}. For the simulation presented in section~\ref{collisions}, impurity collisions~\cite{JOREK_kinetic_Korving_2024_POP} with the background plasma are included, using the Homma operator~\cite{Imp_transport_Homma_2016_NF, Imp_transport_Homma_2013_JCP}. The collision operator captures the classical and the neoclassical Pfirsch--Schlüter transport for the impurity particles. All charge states of the impurities are modelled. At the time when these simulations were run, pumping, molecular reactions, interactions between the deuterium neutrals and the impurities, and self-collisions for either particle species were not yet included in the model.

\begin{figure}[H]
    \centering
    \includegraphics[width=0.8\textwidth]{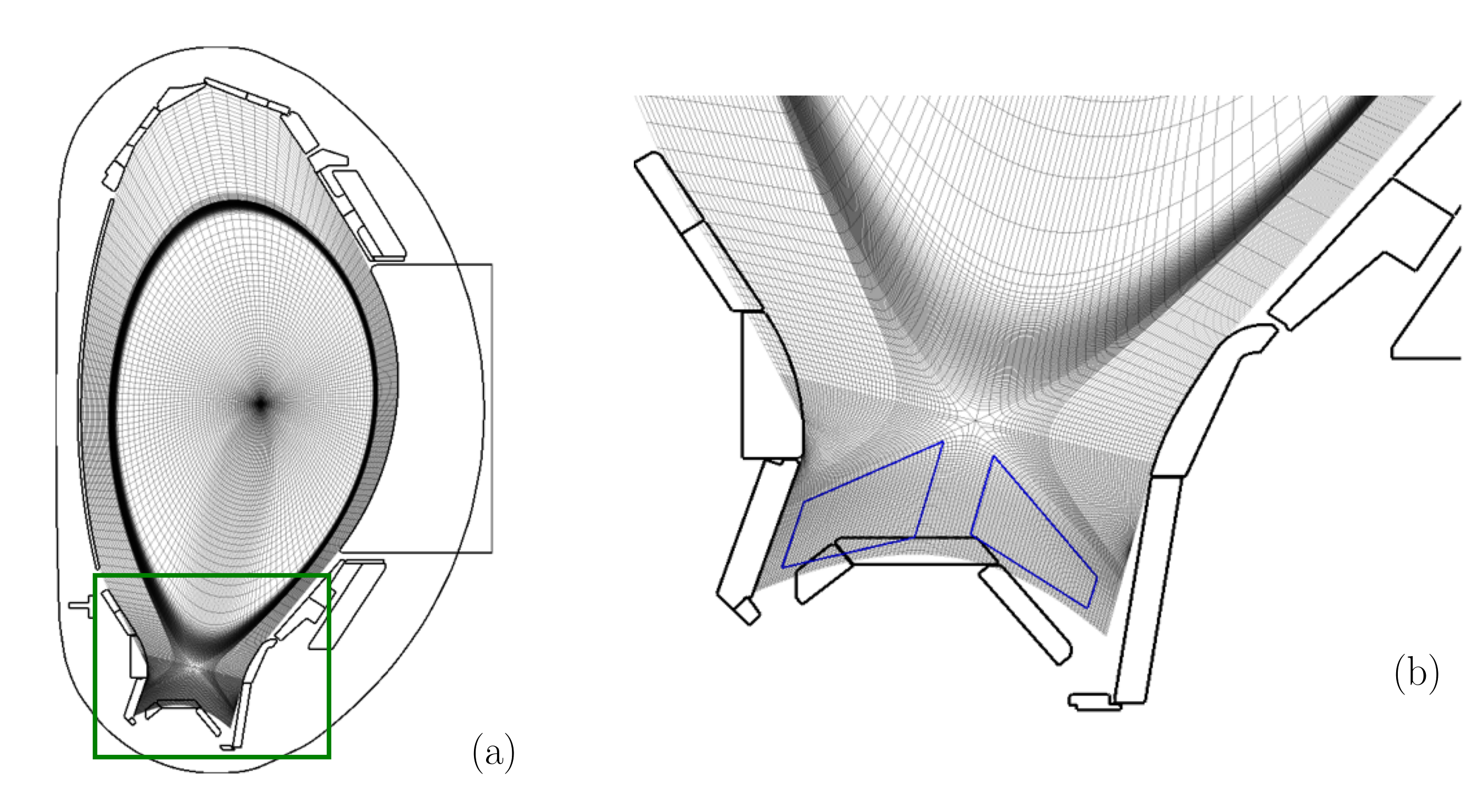}
    \caption{(a) the flux-aligned grid with the grid-to-wall extension to the ASDEX Upgrade first wall. The magnetic equilibrium corresponds to AUG discharge~\#38773 at 3.65 seconds. (b) an enlarged view of the region of interest for the presented XPR simulations, with the blue quadrilaterals marking where the kinetic particles enter the simulation.}
    \label{grid}
\end{figure}

For the spatial discretisation, figure \ref{grid} shows a 2D flux-aligned grid\footnote{Details on the grid construction can be found in Ref.~\cite{JOREK_Czarny_2008_JCP}.} for the lower single-null configuration in AUG discharge~\#38773 at 3.65 seconds. Additionally, the computational grid was fully extended to the AUG first wall, with a small discrepancy at the outer target. The resolution was greatly increased in the PFR, downstream in the scrape-off layer (SOL) and at the plasma edge. This was mainly to combat the constraint of the large flux expansion near the X-point, so that better resolution could be achieved locally near the X-point for the XPR study. The boundary conditions used for this study correspond to a perfectly conducting wall (Dirichlet boundary conditions for the poloidal magnetic field and current density), Bohm (Mach-1) boundary conditions for the parallel velocity, and sheath boundary conditions for density and temperature~\cite{Divertor_Stangeby_2000_SPP}. At the simulation boundary, incident heat and particle flux from the bulk plasma causes recycling, and kinetic deuterium neutrals or impurities reflect with an albedo of 98\;\%. Aside from the boundary reactions, the kinetic particles also enter the simulation via gas puff, from the blue quadrilaterals shown in figure \ref{grid} (b). The large size of the simulated valves is to avoid concentrated impurity content near the boundary when the PFR is cold, which can lead to numerical problems with the low temperature. Different sizes and locations of the simulated valves had been tested inside the PFR, and no difference was shown on whether detachment or XPR formation would occur or not.

\begin{figure}[H]
    \centering
    \includegraphics[width=1\textwidth]{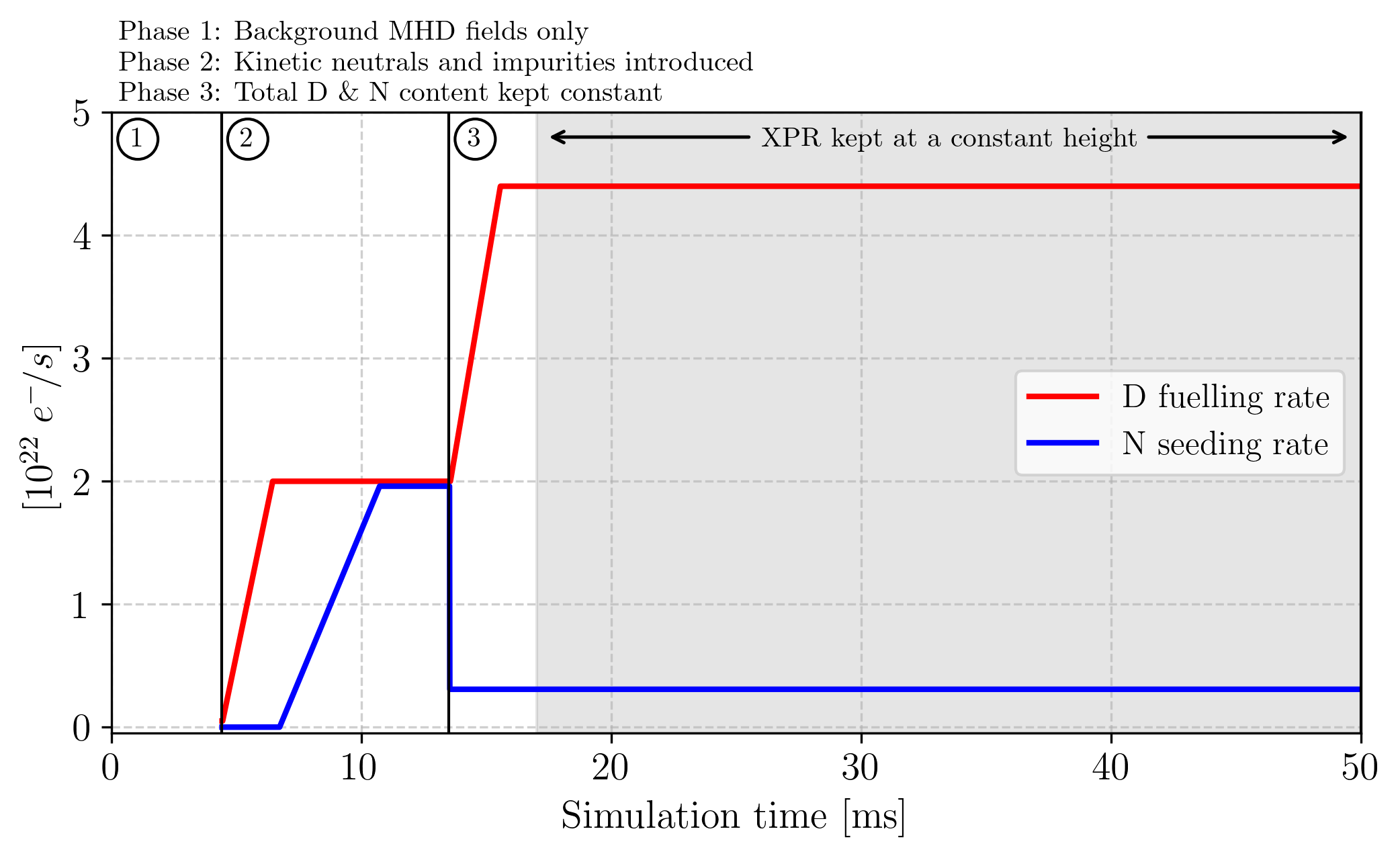}
    \caption{The deuterium fuelling and nitrogen seeding rates through the three simulation phases before reaching the reference time point, with the time window of maintaining a quasi-stationary XPR marked in gray.}
    \label{puffrate}
\end{figure}

The reference simulation set-up comprises three phases, as shown in figure \ref{puffrate}. Phase one (0--4.4\;ms) was run only with the base JOREK model (background deuterium plasma only). At the start of the simulation, experimental plasma density, temperature, and $\mathit{F F'}$ profiles, along with the magnetic equilibrium from the interpretive equilibrium code CLISTE~\cite{CLISTE_Schneider_2000_FED} were used as input, and \textit{ad-hoc} particle and heat diffusion coefficient profiles were used to maintain the plasma profiles. The temperature profile was used to calculate the Spitzer resistivity. During this phase, the Mach-1 boundary conditions~\cite{Divertor_Stangeby_2000_SPP} are propagated inwards from the first wall.

\begin{figure}[H]
    \centering
    \includegraphics[width=1\textwidth]{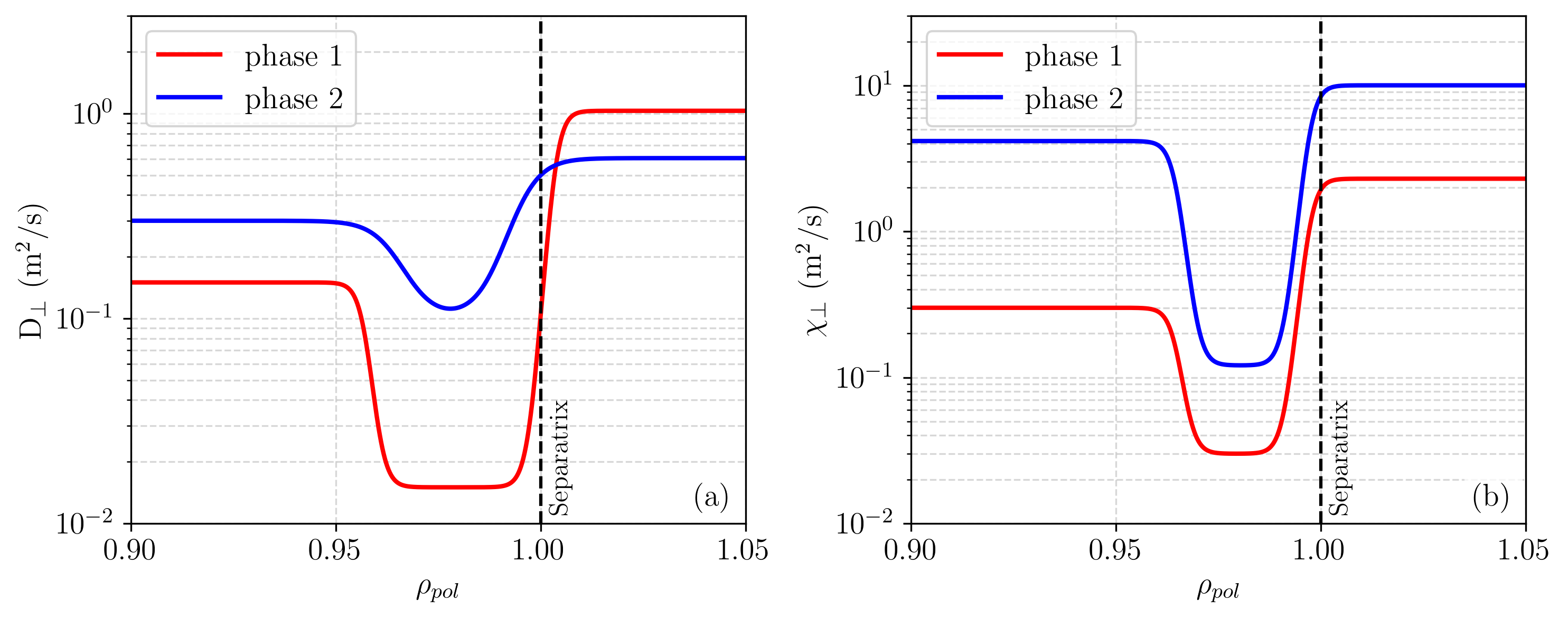}
    \caption{The radial profiles of the (a) particle and (b) heat diffusion coefficients, comparing phase one (background plasma only) and phase two (kinetic particles introduced).}
    \label{diffusion-coeffs}
\end{figure}

In phase two (4.4--13.5\;ms), the kinetic neutrals and impurities were introduced independently. Firstly, the deuterium fuelling rate was ramped up to 2\;$\times$\;10$^{22}$\;e$^-$/s in 2\;ms, with the albedo of 100\;\% initially set for the deuterium neutrals. The outer target cannot be detached with a lower neutral albedo, possibly due to the lack of molecular physics in the model. After the fuelling rate was fully ramped up, nitrogen seeding was initiated and also ramped up to the same rate, with the albedo of 98\;\% set for the impurity particles. It is important to note that, due to the strong particle source from recycling and fuelling in addition to the strong heat sink from all the atomic processes, the diffusion coefficient profiles of the background plasma had to be changed from phase one to phase two\footnote{Ideally, the \textit{ad-hoc} particle and heat diffusion coefficient profiles would be set up to achieve two goals. The first would be to achieve the quasi-stationary XPR, and the second would be to maintain the outer midplane (OMP) profiles to match the experimental measurements. In the presented simulations, only the former was achieved, and the latter remained as an objective of future work.}, in order to reach a steady state after the XPR formation. Figure \ref{diffusion-coeffs} shows the modification of the particle diffusion coefficients $D_{\perp}$ and the heat diffusion coefficients $\chi_{\perp}$. After phase two, the diffusion coefficients were not further modified.

In phase three (13.5--50\;ms), the goal was to keep the fully formed XPR at a constant height. Therefore, the albedo for the neutrals was reduced to 98\;\%, and the fuelling rate was further increased to 4.4\;$\times$\;10$^{22}$\;e$^-$/s. On the other hand, the seeding rate was reduced to 3.1\;$\times$\;10$^{21}$\;e$^-$/s. With these changes, the total (ionised + neutral) deuterium or nitrogen particle content in the simulation saturated at a constant value, and the XPR height remained fixed. Through phase three, a quasi-stationary XPR solution was achieved.

\section{Effects of neutrals and impurities before the XPR formation}\label{formation}

\begin{figure}[H]
    \centering
    \includegraphics[width=1\textwidth]{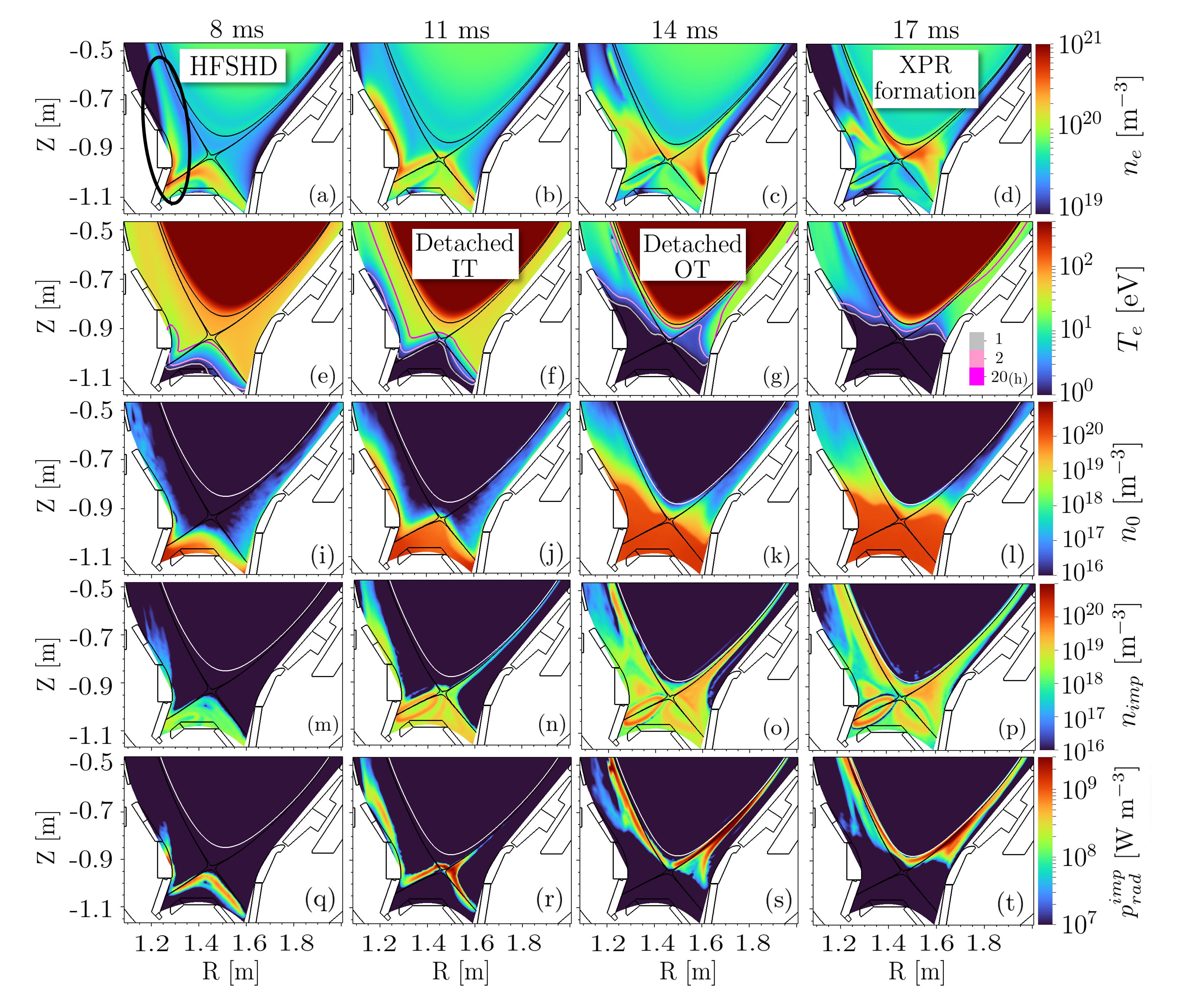}
    \caption{Cross sections of the (a--d) electron density, (e--h) electron temperature with contours at 20, 2 and 1\;eV, (i--l) deuterium neutral density, (m--p) impurity density, and (q--t) impurity radiation density at the time points of the four main events (HFSHD formation, partial detachment, complete detachment and XPR formation).}
    \label{formation-2D}
\end{figure}

\begin{figure}[H]
    \centering
    \includegraphics[width=1\textwidth]{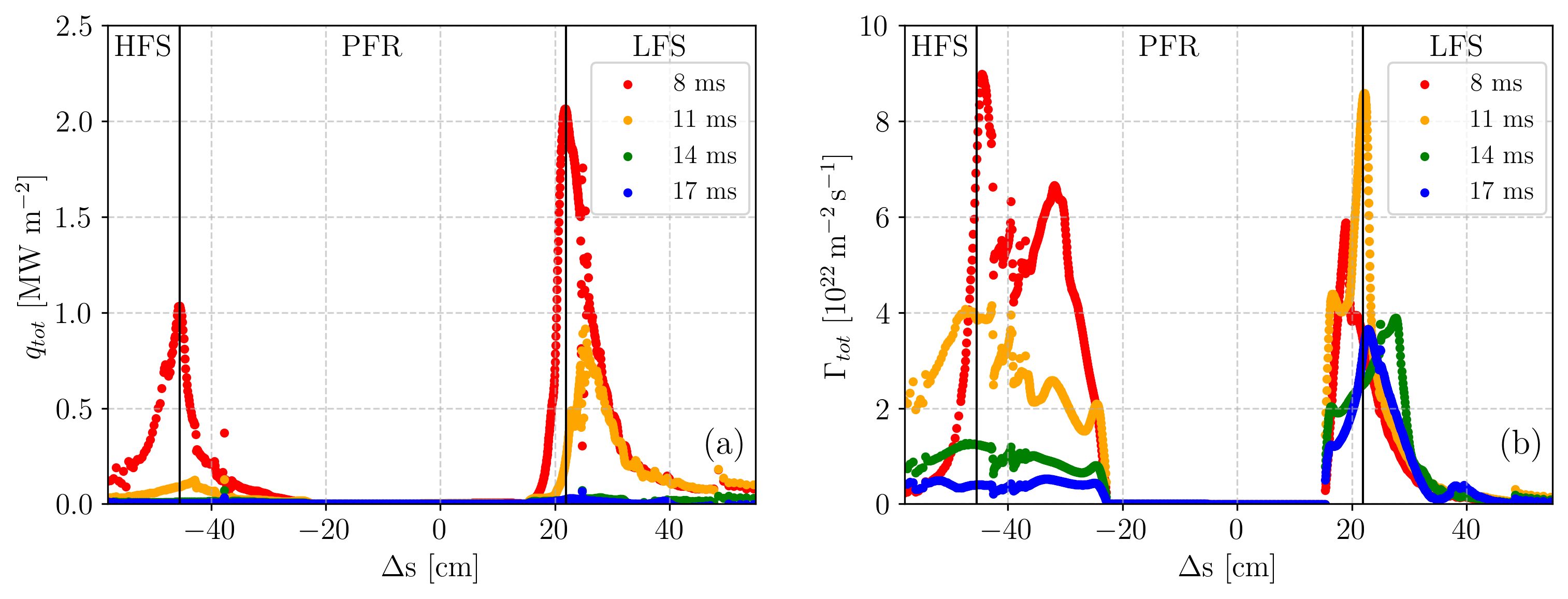}
    \caption{1D profiles of the (a) total heat flux and (b) total particle flux onto the simulation boundary at the same time points as figure \ref{formation-2D}, with the vertical lines marking the inner and outer strike points. Both the heat and particle fluxes are from the background plasma.}
    \label{formation-div}
\end{figure}

Through phase two and the beginning of phase three of the simulation, there are three other events (\textit{high-field-side high-density} (HFSHD) formation~\cite{Detachment_Cavedon_2022_NF}, detached inner target (IT) and detached outer target (OT)~\cite{Detachment_Potzel_2014_NF}) before the stationary XPR is established, each at a different time point. These four time points of the simulation are presented in this section. Figure \ref{formation-2D} shows the cross-sections of the densities of different species, the electron temperature $T_e$, and the impurity line radiation density $p^{imp}_{rad}$. Figure \ref{formation-div} shows the total heat flux $q_{tot}$ and the total particle flux $\Gamma_{tot}$ along the divertor targets. 

In the middle of phase two (8\;ms), the neutral recycling and the high fuelling rate lead to the HFSHD formation, which is shown in figure \ref{formation-2D} (a). The high density\footnote{The electron density shown in this article includes the contribution from both the background deuterium plasma and the free electrons from the (partially) ionized impurities.} plasma volume stretches along the inner target and extends in the near SOL to the inner midplane. (e) shows that the inner target is starting to cool down. Regarding the impurity radiation, (m) and (q) show that the impurities mainly accumulate and radiate in the private flux region, and they begin to leave the PFR along the inner target. At this time point, the ramp-up of the seeding rate has just been started. In figure \ref{formation-div}, $q_{tot}$ and $\Gamma_{tot}$ peak at both strike points, indicating attached inner and outer targets.

Towards the end of phase two (11\;ms), the fully ramped-up seeding and the fuelling lead to a detached inner target. As shown in figure \ref{formation-2D} (f), $T_e$ is reduced to around 1\;eV along the IT and inside the PFR. It is important to note that, for the numerical stability of the simulation, impurities radiation, charge exchange and ionisation are turned off at $T_e$\;$<$\;1\;eV. Recombination is turned off at $T_e$\;$<$\;0.2\;eV. (b) shows that the high density plasma volume stretches along the two strike lines at the downstream SOL, and its upstream extension towards the inner midplane broadens into the farSOL. (n) and (r) show that some impurities are transported into the low-field-side (LFS) SOL and slightly above the X-point, leading to a radiating mantle in these regions. In figure \ref{formation-div}, the peaks of $q_{tot}$ are almost flattened on the inner and halved on the outer target. $\Gamma_{tot}$ is also reduced on the inner target but is more strongly peaked on the outer target.

In the beginning of phase three (14\;ms), the outer target is also detached and a complete detachment is achieved in the simulation. This is shown in figure \ref{formation-div}, where $q_{tot}$ is completely flattened on both targets, and $\Gamma_{tot}$ is also strongly reduced on the inner target. In figure \ref{formation-2D} (k) and (o), the neutrals are well compressed at the downstream SOL but are able to enter the confined region where the flux expansion is large, and a significant amount of impurities are able to enter there as well. In the region between the contours of $T_e$\;$=$\;20 and 2\;eV in (g), the impurities can strongly radiate, meaning the XPR is forming, as shown in (s). Additionally, (c) shows that upon complete detachment, HFSHD no longer extends to the inner midplane.

Shortly after the saturation of the deuterium and the nitrogen content (17\;ms), the simulation proceeds towards a stable solution, and a dense plasma volume is formed between the separatrix and the flux surface of $\rho_{pol}$\;$=$\;0.994 which overlaps with the $T_e$\;$=$\;20\;eV contour downstream, as shown in figure \ref{formation-2D} (d) and (h). The density plot also shows the loss of the HFSHD, which was experimentally observed in Ref.~\cite{Detachment_Cavedon_2022_NF}. (l) and (p) show that neutrals and impurities are accumulated in the XPR volume, whilst a small amount of impurities even transported above the XPR. (t) then shows that the XPR is formed along this flux tube, and the radiating mantle is mainly in the confined region, with the LFS SOL also strongly radiating. Figure \ref{formation-div} shows that the $q_{tot}$ and $\Gamma_{tot}$ profiles are similar to the previous time point, meaning the detachment state does not change.

\section{A quasi-stationary XPR solution}\label{ref}

Through phase three of the simulation, after the fuelling and seeding rates are adjusted so that the total deuterium and nitrogen content remains constant, the simulation continues to run for more than 30 ms without any changes. As a result, the XPR height also remains constant, confirming that the XPR solution is quasi-stationary. This section presents the last time point of phase three (50\;ms), which is also used as the reference for the next two sections, where the seeding rate will be changed to move the XPR. This reference time point is hereby denoted as the "stationary case".

\begin{figure}[H]
    \centering
    \includegraphics[width=1\textwidth]{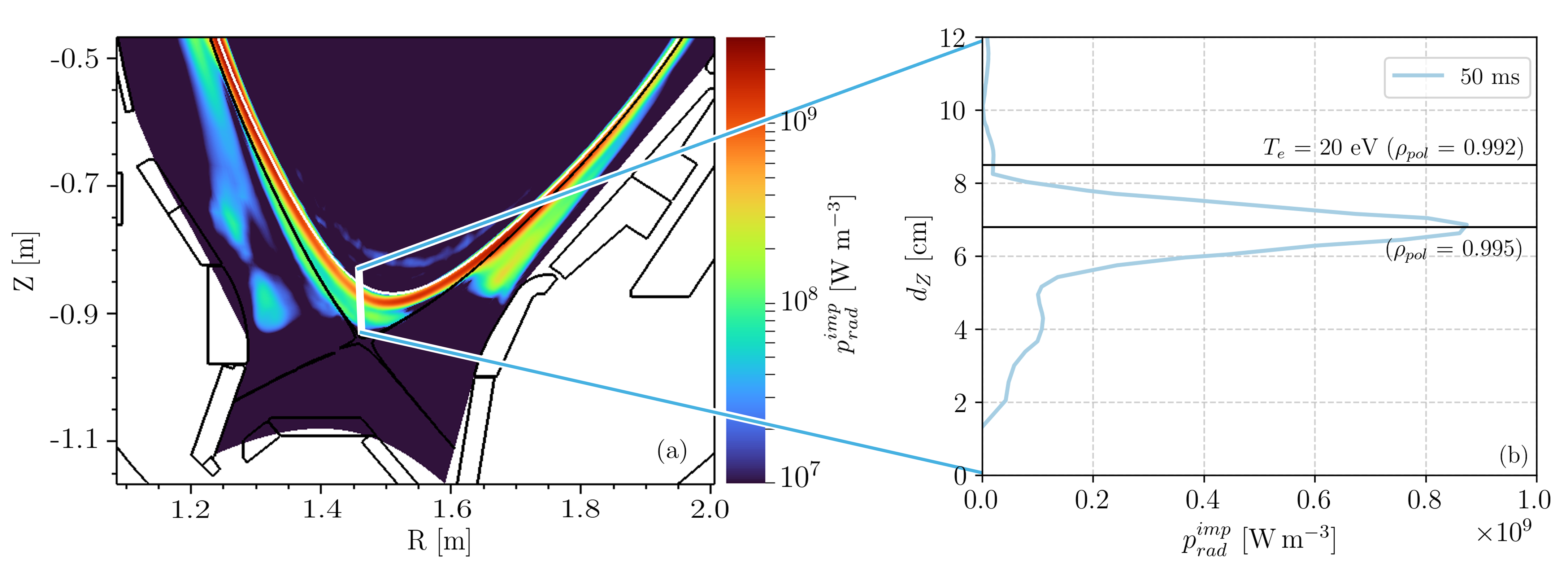}
    \caption{(a) cross section of the impurity radiation density and (b) the vertical profile above the X-point in the stationary case, with the horizontal lines marking the height of the radiation peak and the upper bound of the radiating mantle.}
    \label{ref-Prad}
\end{figure}

Figure \ref{ref-Prad} (a) shows a cross-sectional plot of the impurity radiation density $p^{imp}_{rad}$ for the stationary case. Before further discussion, it is important to note that the impurity collisions~\cite{JOREK_kinetic_Korving_2024_POP} are not yet included in this presented simulation, as mentioned in section \ref{setup}. Therefore, the impurity particles can be freely transported via $E$$\times$$B$ drift without experiencing friction from the background plasma, and the radiating mantle appears poloidally elongated along the residing flux tube. The inclusion of the impurity-background collisions, which is shown to prevent the poloidal elongation of the XPR, is further discussed in section \ref{collisions}.

Figure \ref{ref-Prad} (b) shows a vertical cut of $p^{imp}_{rad}$ above the X-point. In the following discussion, the vertical distance between the X-point and the peak of $p^{imp}_{rad}$ is defined as the XPR height. Above the radiation peak, the radiating mantle extends to an upper bound, where the local electron temperature reaches 20\;eV, as it is the temperature above which the nitrogen particles no longer strongly radiate. Lastly, the XPR core is defined as the cold and dense plasma volume between the radiating mantle and the separatrix. In the stationary case, the XPR height remains constant at around 6.8\;cm ($\rho_{pol}$\;=\;0.995), with the upper bound being at 8.5\;cm ($\rho_{pol}$\;=\;0.992). The total radiative fraction $f_{rad}$\;$=$\;$P_{rad,tot}/P_{heat}$ reaches 51\;\%, with the radiative power mainly contributed from the impurities and the total heating power fixed at 10\;MW.

\begin{figure}[H]
    \centering
    \includegraphics[width=1\textwidth]{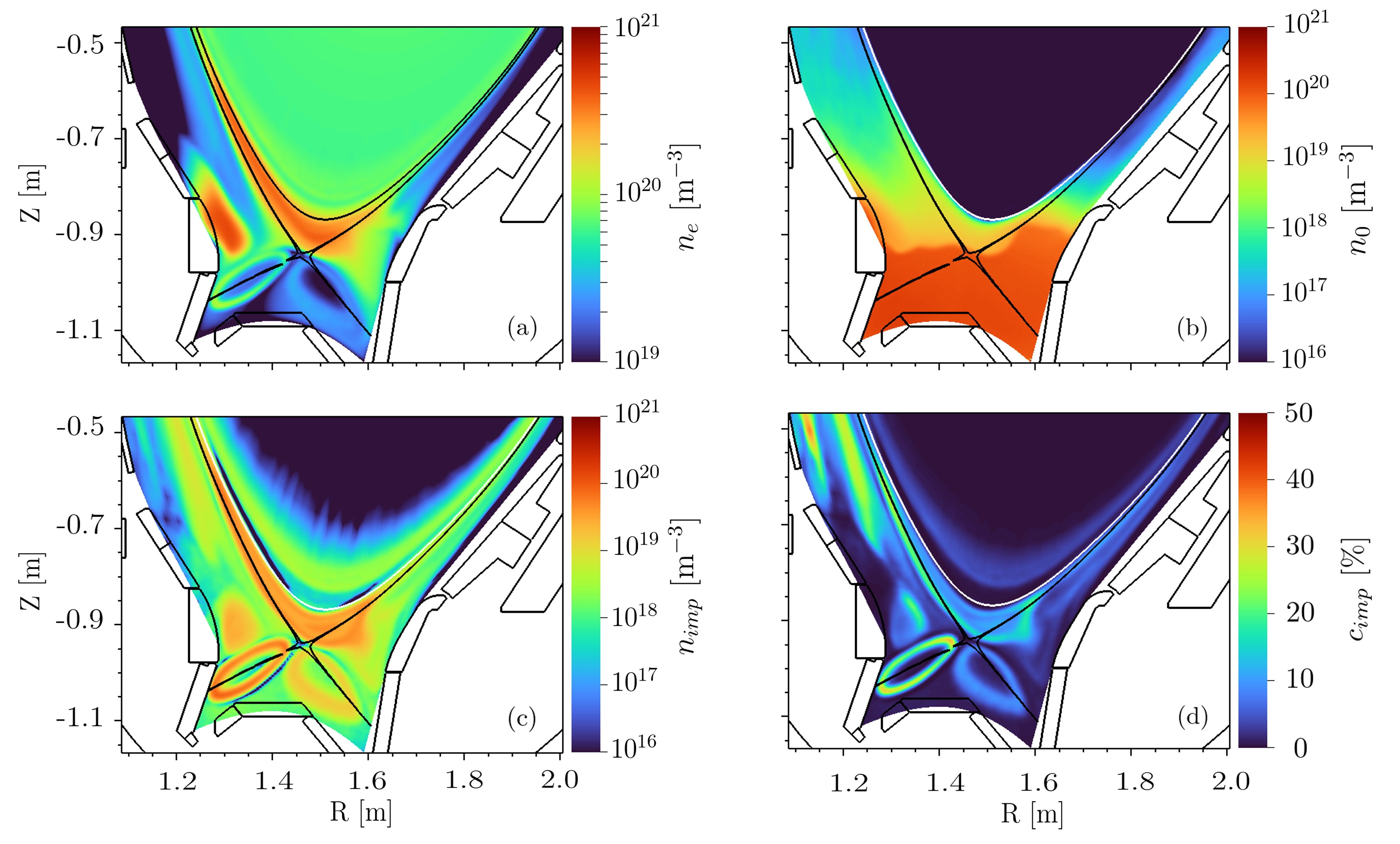}
    \caption{Cross sections of the (a) total electron density, (b) neutral density, (c) impurity ion density, and (d) impurity concentration in the stationary case. The impurity concentration is calculated as the ratio between the impurity ion density and the total electron density.}
    \label{ref-densities}
\end{figure}

Figure \ref{ref-densities} shows the densities of various species (a: electrons, b: deuterium neutrals, c: impurity ions) in the stationary case. The electron density in the XPR core is around 3\;$\times$\;10$^{20}$\;m$^{-3}$, around 4 times the core density. This again includes both the background deuterium plasma and the impurity contribution. The neutral density is around one order of magnitude lower than the electron density in the XPR core. At the upper bound of the radiating mantle, $n_0$ reduces to around 1\;$\times$\;10$^{17}$\;m$^{-3}$. Impurities are mainly concentrated in the XPR core and the radiating mantle, with the density also around one order of magnitude lower than the electron density. Further inside the confined region, impurities can also be transported there due to the lack of impurity-background collisions. However, this does not lead to any further movement of the XPR, as a sufficient number of neutrals is required to initiate the formation of the new XPR volume, shifting the radiating mantle up and expanding the XPR core~\cite{XPR_model_Stroth_2022_NF}. (d) shows a cross section of $c_{imp} = n_{imp}/n_e$. Inside the separatrix, the XPR core shows to have a high impurity concentration of 10--15\;\%. This is similar to the experimental results in AUG~\cite{XPR_Henderson_2020_IAEA} and in the SOLPS-ITER simulations in TCV~\cite{SOLPS_Sun_2025_arXiv}. Outside the separatrix, $c_{imp}$ can also be high in some regions, mainly due to the low temperature and consequently the low plasma density, as the plasma is completely detached.

\begin{figure}[H]
    \centering
    \includegraphics[width=1\textwidth]{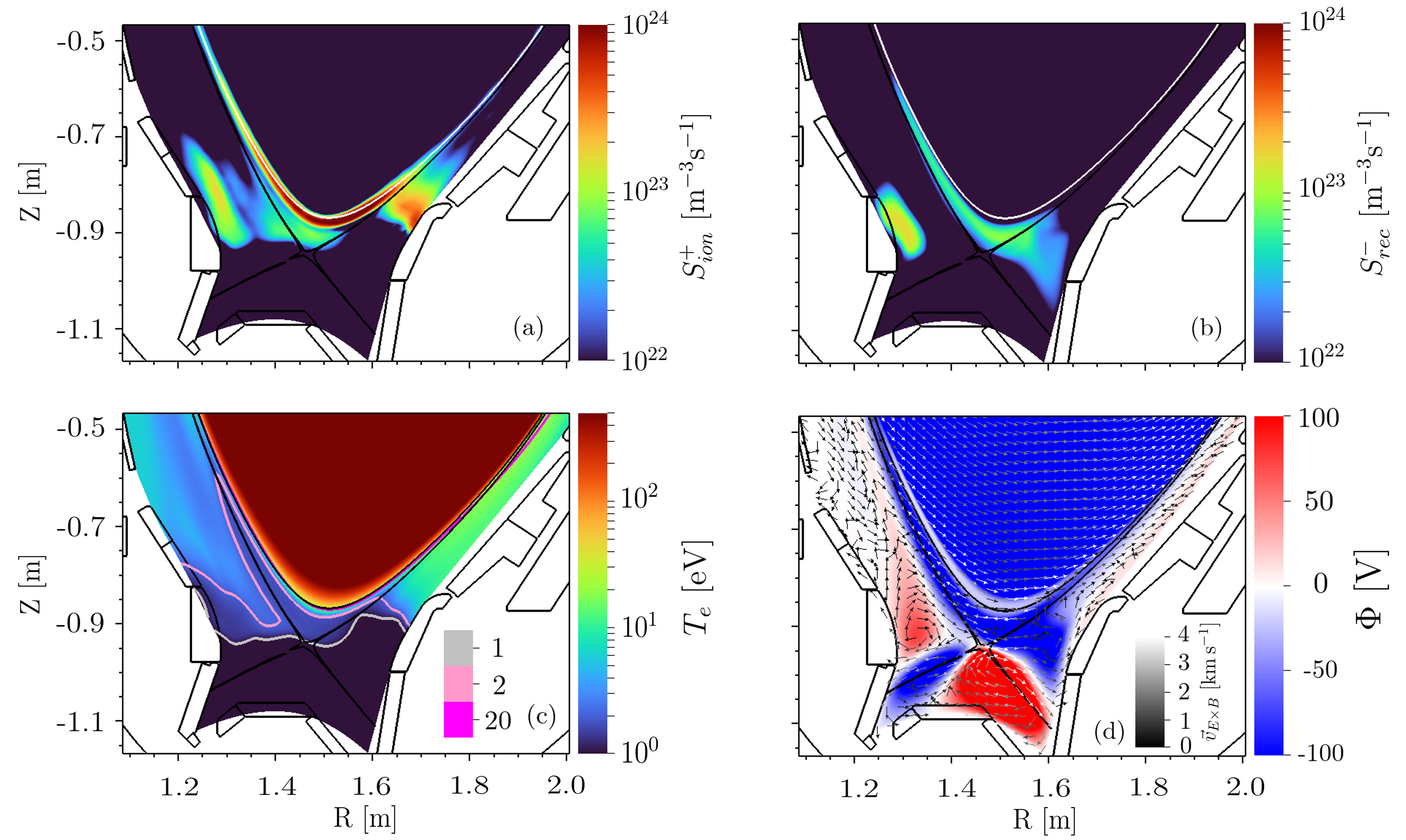}
    \caption{Cross sections of the (a) ionisation rate, (b) recombination rate, (c) electron temperature with the contours at 1, 2, and 20\;eV, and (d) electrostatic potential overlapped with the $E$$\times$$B$ drift velocity in the stationary case.}
    \label{ref-others}
\end{figure}

Along with the impurity radiation, the two other crucial physical phenomena in an XPR are the ionisation and recombination of the deuterium species~\cite{XPR_model_Stroth_2022_NF, SOLPS_Pan_2023_NF}. Figure \ref{ref-others} (a) shows that the ionisation rate $S^{+}_{ion}$ is most profound in the radiating mantle. The peak of $S^{+}_{ion}$ locates at 7.5\;cm, slightly higher than the peak of $p^{imp}_{rad}$, and the whole ionisation front reaches slightly above the radiating mantle. This indicates that neutrals can penetrate through the XPR core and the radiating mantle, reaching further inside the confined region, and the ionisation front contributes to reducing the local $T_e$ to around 20\;eV. On the other hand, (b) shows that the deuterium recombination mainly takes place in the XPR core.

Figure \ref{ref-others} (c) shows a strong gradient of the electron temperature from the radiation peak (2.8\;eV) to the upper bound of the radiating mantle (20\;eV). Over this region, the impurity radiation and the ionisation of the neutrals are the major heat sinks of the incoming power from the plasma core. It is important to note that $T_e = T_i$ is assumed, where the impurity radiation would only reduce $T_e$, and this would lower the height of the ionisation front. In the XPR core, both effects along with the deuterium recombination take place, and $T_e$ is maintained around 1 to 2\;eV.

Regarding the $E$$\times$$B$ transport of the impurities, figure \ref{ref-others} (d) shows the electrostatic potential. Above the radiating mantle, a potential hill is present. This is similar to the SOLPS-ITER XPR simulation, as shown by figure 11(f) in Ref.~\cite{SOLPS_Pan_2023_NF}. The resulting radial electric field on the HFS leads to a poloidal $E_r$$\times$$B$ drift from the XPR towards the inner midplane. This is then responsible for the poloidal extension of the high density plasma volume and the impurity transport, towards the HFS. In the recent simulations with the turbulence code GRILLIX~\cite{GRILLIX_Eder_2025_NF}, a similar $E$$\times$$B$ vortex above the XPR is also observed. As a side note, the potential well and hill in the PFR lead to some impurities drifting around them, and the impurity density becomes comparable to that inside the XPR. Although these structures look peculiar, they are outside the region of interest for this time point of the simulation, and the local $T_e$ is too low for the accumulated impurities to strongly affect the background plasma via radiation. Therefore, these structures are noted but not further discussed.

To sum up, in the radiating mantle, the ionisation of the neutrals serves as an ion source for the XPR, and it is also the main heat sink responsible for reducing $T_e$ to around 20 eV, low enough for the impurities to strongly radiate. Towards the radiation peak, the impurity radiation becomes gradually dominant and further reduces $T_e$ to 2.8\;eV. In the XPR core, deuterium recombination occurs as a source of neutrals, along with the neutrals from the fuelling gas puff, and the neutrals can penetrate upwards through the radiating mantle to fuel the ionisation front. Additionally, a radial electric field is created above the XPR upon its formation, and the resulting $E_r$\;$\times$\;$B$ drift leads to a poloidal transport for the deuterium plasma and the impurities towards the inner midplane.


\section{Development into a MARFE}\label{marfe}

To dynamically evolve the XPR, the stationary case branches out from the 50\;ms time point to several variations, each with a different seeding rate, without changing any other setting. Since the recycling coefficients are unchanged, the total nitrogen content gradually increases with a higher seeding rate, and vice versa. In the following two sections, two of these cases are shown to demonstrate how the change of the nitrogen content affects the XPR. In this section, the MARFE case is presented, where the seeding rate is increased by a factor of five, to 1.54\;$\times$\;10$^{22}$\;e$^-$/s. During this simulation, the XPR develops vertically upwards and eventually becomes MARFE-like before the simulation crashes due to numerical issues with the extremely low temperature in the MARFE core. The four time points chosen to present this development are as follow:

\begin{itemize}
    \item 50\;ms serves as the reference.
    \item At 58\;ms, the XPR starts to develop vertically upwards.
    \item At 61\;ms, $f_{rad}$ reaches 60\;\%. The electron temperature slightly beneath the radiation peak starts to further decrease and volumetric recombination is enhanced locally. The coldest part of the XPR core starts to become MARFE-like.
    \item At 64\;ms, a MARFE-like scenario is achieved.
\end{itemize}

\begin{figure}[H] 
    \centering
    \includegraphics[width=1\textwidth]{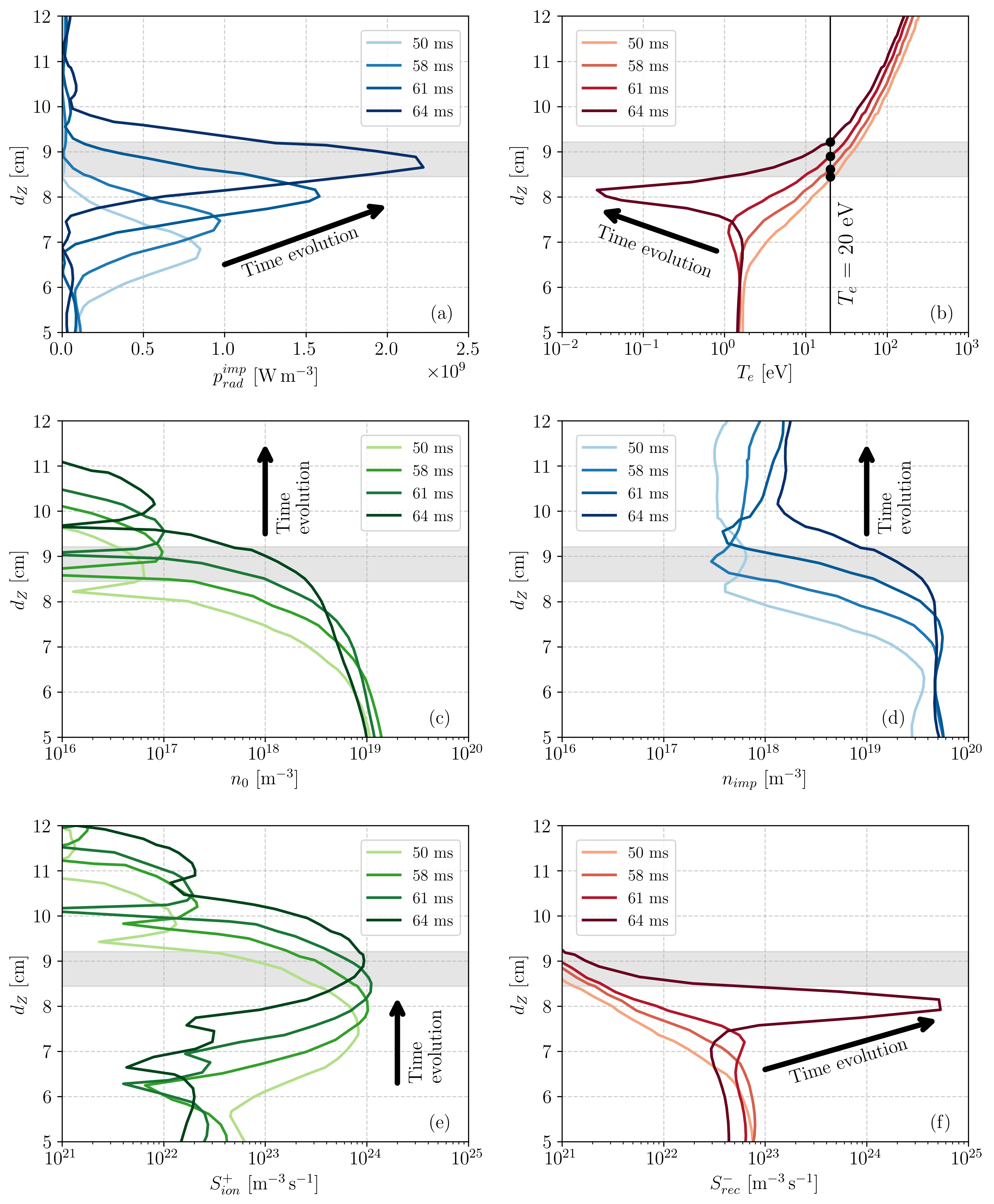}
    \caption{Vertical profiles of the (a) impurity radiation density, (b) electron temperature, (c) neutral density, (d) impurity density, (e) ionisation rate, and (f) recombination rate during the MARFE case. The gray band marks the upward shift of the position where $T_e$\;=\;20\;eV at each time point.}
    \label{marfe-vertical}
\end{figure}

Figure \ref{marfe-vertical} shows the vertical profiles of various quantities over the four time points. (a) shows the peak intensity of $p^{imp}_{rad}$ growing over time, and the XPR height shifts from 6.8 to 8.7\;cm. (b) shows the development of temperature in the XPR, and the position where $T_e$\;=\;20\;eV shifts upwards. Between the last two time points, $T_e$ starts to drastically decrease to well below 1\;eV below the radiation peak, indicating a transition to a MARFE. Additionally, it is noticed that, for a MARFE case, the vertical position of $T_e$\;=\;20\;eV does not serves as a good estimate for the upper bound of the radiating mantle.

Figure \ref{marfe-vertical} (c) and (d) show the number density of neutrals and impurities, respectively. For the neutrals, the distribution simply shifts vertically upwards. For the impurities, the vertical shift is accompanied by an increased density further into the plasma core, above the XPR/MARFE.

Figure \ref{marfe-vertical} (e) and (f) show the rates of ionisation and recombination, respectively. In the reference time point, the ionisation front lays slightly above the radiating mantle, whilst recombination is most profound below it. In contrast, after the MARFE formation, the radiating mantle becomes more overlapped with the ionisation front, and the MARFE core slightly below the radiation peak strongly recombines due to the low temperature and high local plasma density. This increase of the recombination rate during the transition to a MARFE is also described in Refs.~\cite{XPR_model_Stroth_2022_NF, MARFE_model_Simakov_2000_PoP}.

\begin{figure}[H] 
    \centering
    \includegraphics[width=1\textwidth]{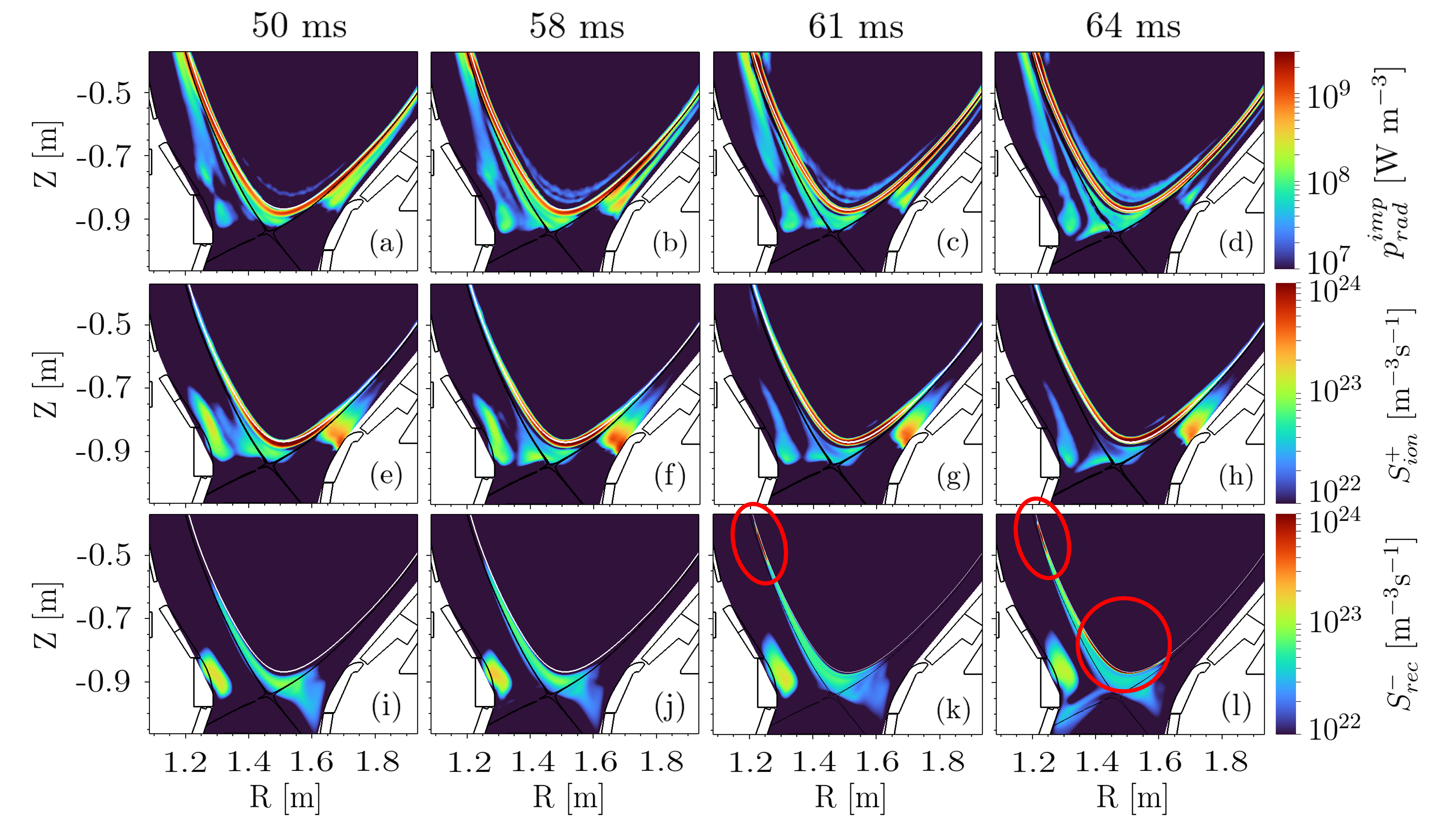}
    \caption{Cross sections of the (a--d) impurity radiation density, (e--h) ionisation rate, and (i--l) recombination rate during the MARFE case. The flux surface ($\rho_{pol}$\;=\;0.992) which overlaps with the upper bound of the radiating mantle at 50\;ms is marked in white in all the figures.}
    \label{marfe-2D}
\end{figure}

Figure \ref{marfe-2D} shows the cross sections of $p^{imp}_{rad}$, $S^{+}_{ion}$ and $S^{-}_{rec}$ at the same time points as figure \ref{marfe-vertical}. To better present the vertical movement of the radiating mantle and the ionisation front, the flux surface ($\rho_{pol}$\;=\;0.992), which marks the upper bound of the radiating mantle at the reference time point, is also plotted in white. From (a) to (d) and from (e) to (h), the radiating mantle and the ionisation front simply shift upwards. From (i) to (l), the region with significant recombination only extends slightly upwards. However, at 61\;ms, a small plasma volume with high $S^{-}_{rec}$ and $T_e$ well below 1\;eV appears near the inner midplane, which is recognised as a MARFE in the context of this article. The formation of the MARFE near the inner midplane is due to the poloidal $E_r$$\times$$B$ drift, which leads to high background plasma density and high impurity content in that region. At 64\;ms, another MARFE volume is present above the X-point, which is near the region where the radiating mantle and the ionisation front overlap. This is also shown in figure \ref{marfe-vertical}, and the discussion in this section is focused on this MARFE volume.

To sum up, the process of an XPR moving upwards and turning into a MARFE can be summarized into a sequence of the following effects:

\begin{itemize}
    \item The increase of $n_{imp}$ in the radiating mantle leads to a shifts of the radiation peak.
    \item The radiative mantle starts to overlap with the ionisation front. Locally, the impurity radiation becomes the dominating heat sink.
    \item The ionisation front develops further into the plasma core.
    \item A new XPR solution with a larger XPR height is reached.
    \item As $p^{imp}_{rad}$ becomes more strongly peaked, the overlap of the radiating mantle and the ionisation front no longer simply causes the XPR to expand upwards. Simultaneously, the dense plasma volume below the radiation peak eventually reaches a temperature low enough for it to strongly recombine. The strong recombination reduces the plasma density, and $T_e$ is further reduced to well below 1\;eV.
    \item The extraordinarily low temperature and the high recombination rate in this plasma volume suggests the formation of a MARFE.
\end{itemize}

\section{Retreat and the gradual loss of the XPR}\label{retreat}

In contrast to the previous section, this section presents the retreating case, where the seeding rate is reduced to 0. During this simulation, 25\;ms after the seeding is turned off, the XPR starts to develop vertically downwards and the radiating mantle eventually moves outside the confined region, meaning the XPR is lost. The four time points chosen to present this development are as follow:

\begin{itemize}
    \item 50\;ms serves as the reference.
    \item At 75\;ms, the XPR starts to develop vertically downwards.
    \item At 84\;ms, the radiating mantle start to move outside of the confined region and enter the SOL.
    \item At 91\;ms, the XPR is entirely lost.
\end{itemize}

\begin{figure}[H] 
    \centering
    \includegraphics[width=1\textwidth]{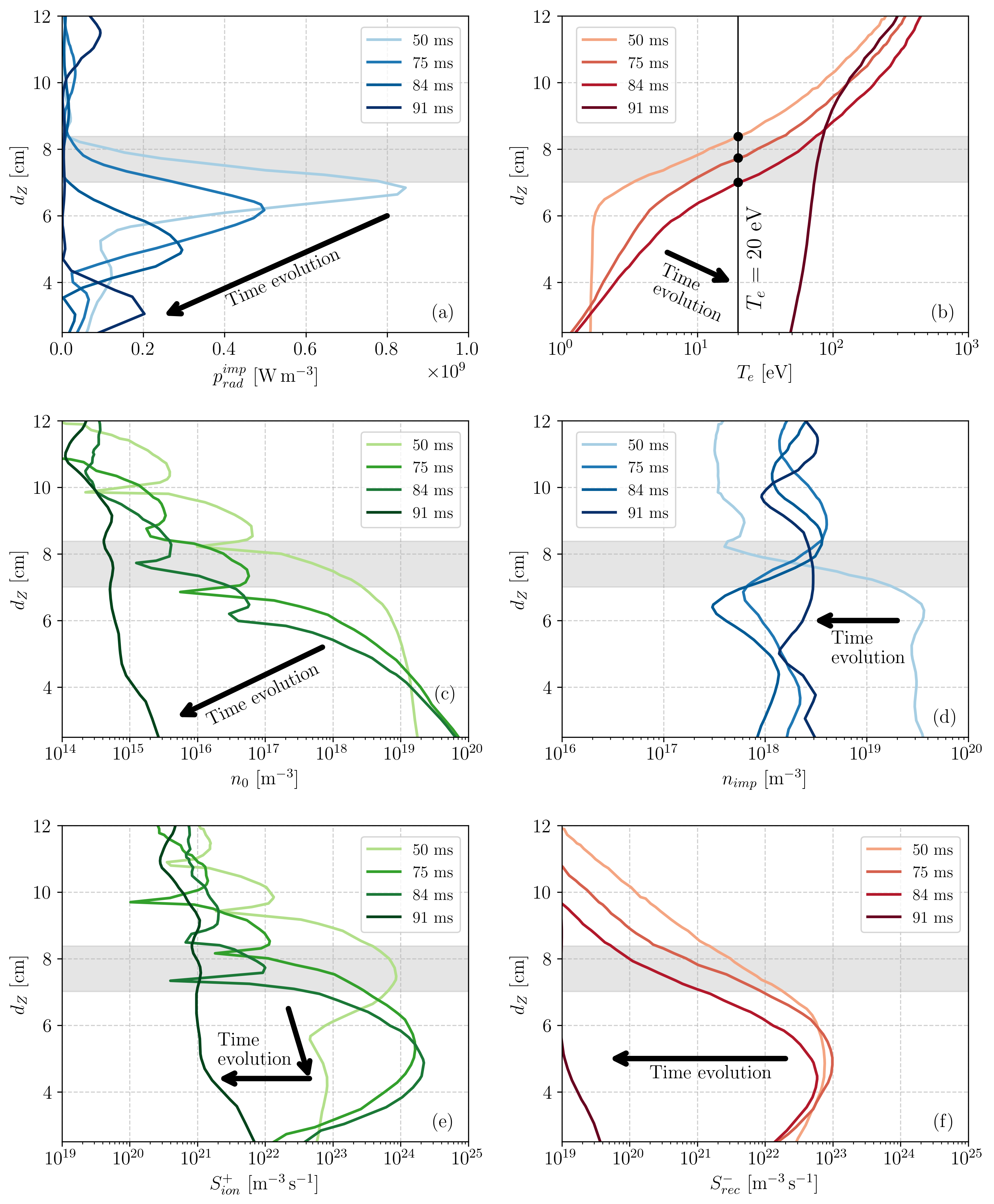}
    \caption{Vertical profiles of the (a) impurity radiation density, (b) electron temperature, (c) neutral density, (d) impurity density, (e) ionisation rate, and (f) recombination rate during the retreating case. The gray band marks the downward shift of the position where $T_e$\;=\;20\;eV at the first three time points. At the last time point, where the XPR is lost, $T_e$ is entirely above 20\;eV in the confined region.}
    \label{retreat-vertical}
\end{figure}

Figure \ref{retreat-vertical} shows the vertical profiles of various quantities over the four time points. (a) shows the peak intensity of $p^{imp}_{rad}$ and the XPR height reducing over time. (b) shows that the XPR core gradually heats up, and the cold plasma volume ($T_e$\;$<$\;20\;eV) shrinks. At the last time point, $T_e$ is entirely above 20\;eV in the confined region, indicating the lost of the XPR. Additionally, the outer target is again attached.

Figure \ref{retreat-vertical} (c) and (d) show the number density of neutrals and impurities, respectively. For the neutrals, the distribution first shifts vertically downwards, and then almost all the neutrals in the confined region are ionised away between the last two time points. For the impurities, there is no vertical shift of the distribution. Instead, $n_{imp}$ is first strongly reduced in the XPR core and the radiating mantle, which is responsible for the weakening of $p^{imp}_{rad}$, and then the distribution saturates, even after the loss of the XPR. It is important to note that, between the first two time points, $n_{imp}$ slightly increases above the XPR. This is caused by the changes on the poloidally elongated potential hill above the XPR, and the resulting $E$$\times$$B$ drift transports some of the impurities from the XPR core further into the confined region.

Figure \ref{retreat-vertical} (e) and (f) show the rates of ionisation and recombination, respectively. In contrast to the stationary case, as the XPR retreats, the ionisation front moves from slightly above the radiation peak to more overlapping with it, and the peak ionisation rate is increased by a factor of two. This indicates that, in the XPR core, the weakened power dissipation via impurity radiation leads to a temperature increase, which results in a higher ionisation rate of the neutrals. Eventually, the XPR is lost when the neutrals are depleted. Upon losing the XPR, both $S^{+}_{ion}$ and $S^{-}_{rec}$ are strongly reduced.

\begin{figure}[H] 
    \centering
    \includegraphics[width=1\textwidth]{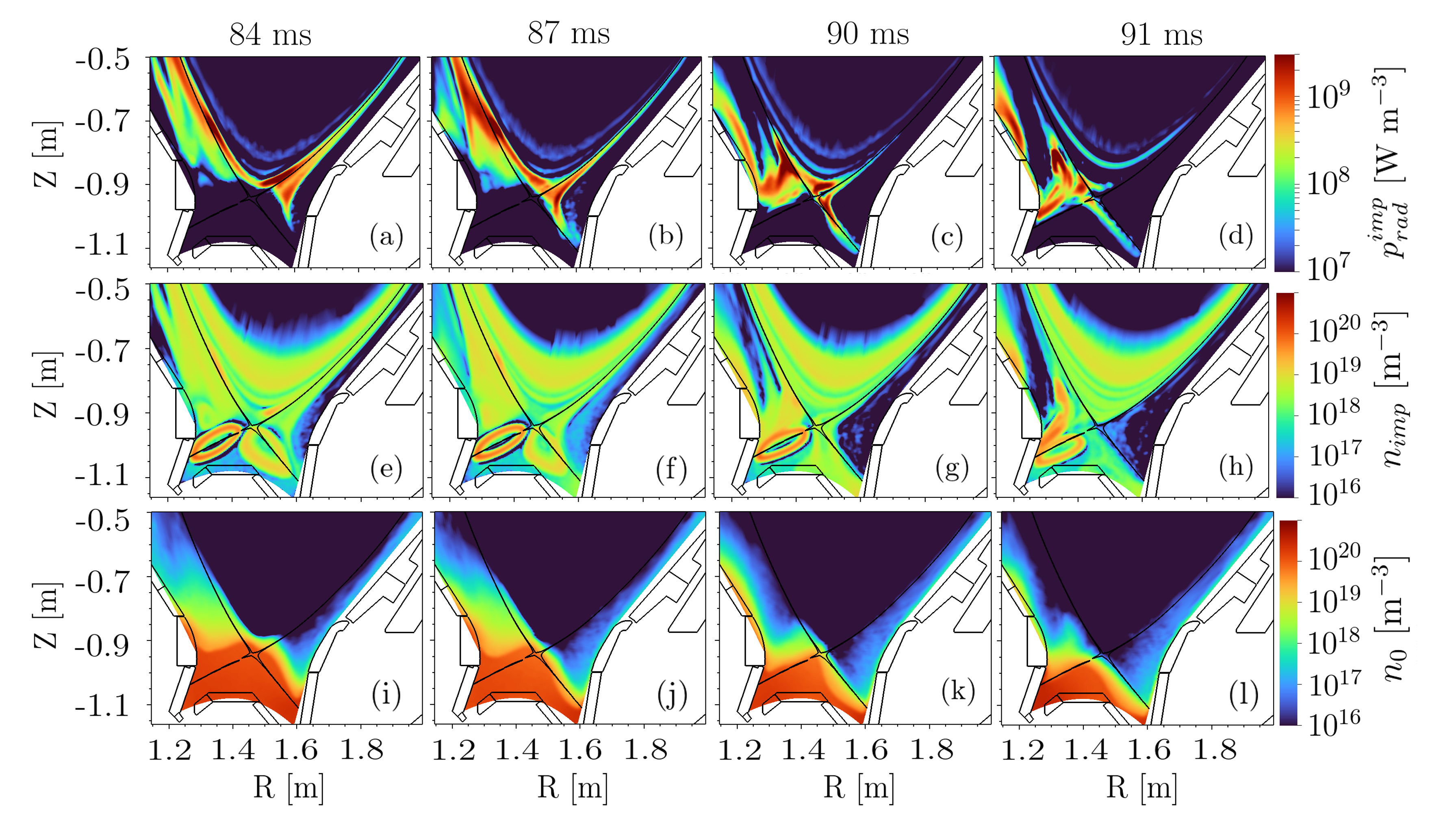}
    \caption{Cross sections of the (a--d) impurity radiation density, (e--h) impurity density, and (i--l) deuterium neutral density between the last two time points in the retreating case.}
    \label{retreat-2D}
\end{figure}

Figure \ref{retreat-2D} shows the cross sections of $p^{imp}_{rad}$, $n_{imp}$, and $n_{0}$ during the loss of the XPR between the last two time points. From (a) to (d), the radiating mantle gradually moves outside the confined region and enters the SOL. At the last time point, the radiating mantle locates mainly at the downstream HFS SOL, with some radiation still present in the confined region. From (e) to (h), $n_{imp}$ is strongly reduced in the LFS SOL, and the impurities in the SOL are compressed near the inner target. The same compression for the neutrals can be seen from (i) to (l). It is important to note that the radiation in the confined region is caused by the remaining impurities, which are transported there via $E$$\times$$B$ drift when the potential hill above the XPR is still present. After the loss of the XPR, the potential hill is also lost, and the cross-field $E$$\times$$B$ transport in the confined region is weakened. This leads to the impurities being effectively trapped in the confined region, if no other transport mechanism is included in the simulation.

To sum up, this section provide a retrogressive view of the XPR formation, and the process of a XPR moving downwards and eventually being lost can be summarized into a sequence of the following effects:

\begin{itemize}
    \item The reduction of $n_{imp}$ leads to a reduction of power dissipation via $p^{imp}_{rad}$ in the radiating mantle.
    \item The ionisation rate of neutrals increases in the radiating mantle as the temperature rises.
    \item Less neutrals can cross the radiating mantle, leading to the retreat of the ionisation front and consequently the retreat of the XPR.
    \item Upon consuming all the neutrals in the gradually shrinking XPR core, the power balance cannot be maintained and the XPR solution is lost. The radiating mantle moves outside the confined region and remains at the downstream HFS SOL, where the impurities and neutrals in the SOL are compressed.
\end{itemize}

Comparing to the MARFE case, the retreating case shows hysteresis, namely the XPR core shrinks at a lower $n_{imp}$ than when it is created. Although both developments start with a change in $n_{imp}$, the ionisation of the neutrals still sustains the XPR in the retreating case. Only when the neutral content in the XPR core is depleted, the XPR is lost. Therefore, an option for future work is to test whether another quasi-stationary XPR solution can be reached with a reduced seeding rate but an increased fuelling rate.

\section{Effect of the impurity collisions}\label{collisions}

In the simulations shown so far, the only transport mechanism for the impurity particles is the $E$$\times$$B$ drift from the background electric field. This section presents the collisional case, where the collision operator~\cite{Imp_transport_Homma_2016_NF, Imp_transport_Homma_2013_JCP} for the impurity particles is turned on. This transport mechanism is a collective effect from calculating the small-angle Coulomb collisions between the impurity particles and the background plasma, with a much finer time resolution (0.4 ns) than the background MHD simulation time step (2 $\mu$s) and the kinetic particle time step (20 ns). For the collision operator, the velocity distribution of the background plasma is assumed to be a Maxwellian shifted and disturbed by the local plasma flow and heat flux~\cite{JOREK_kinetic_Korving_2024_POP}. 

\begin{figure}[H]
    \centering
    \includegraphics[width=0.7\textwidth]{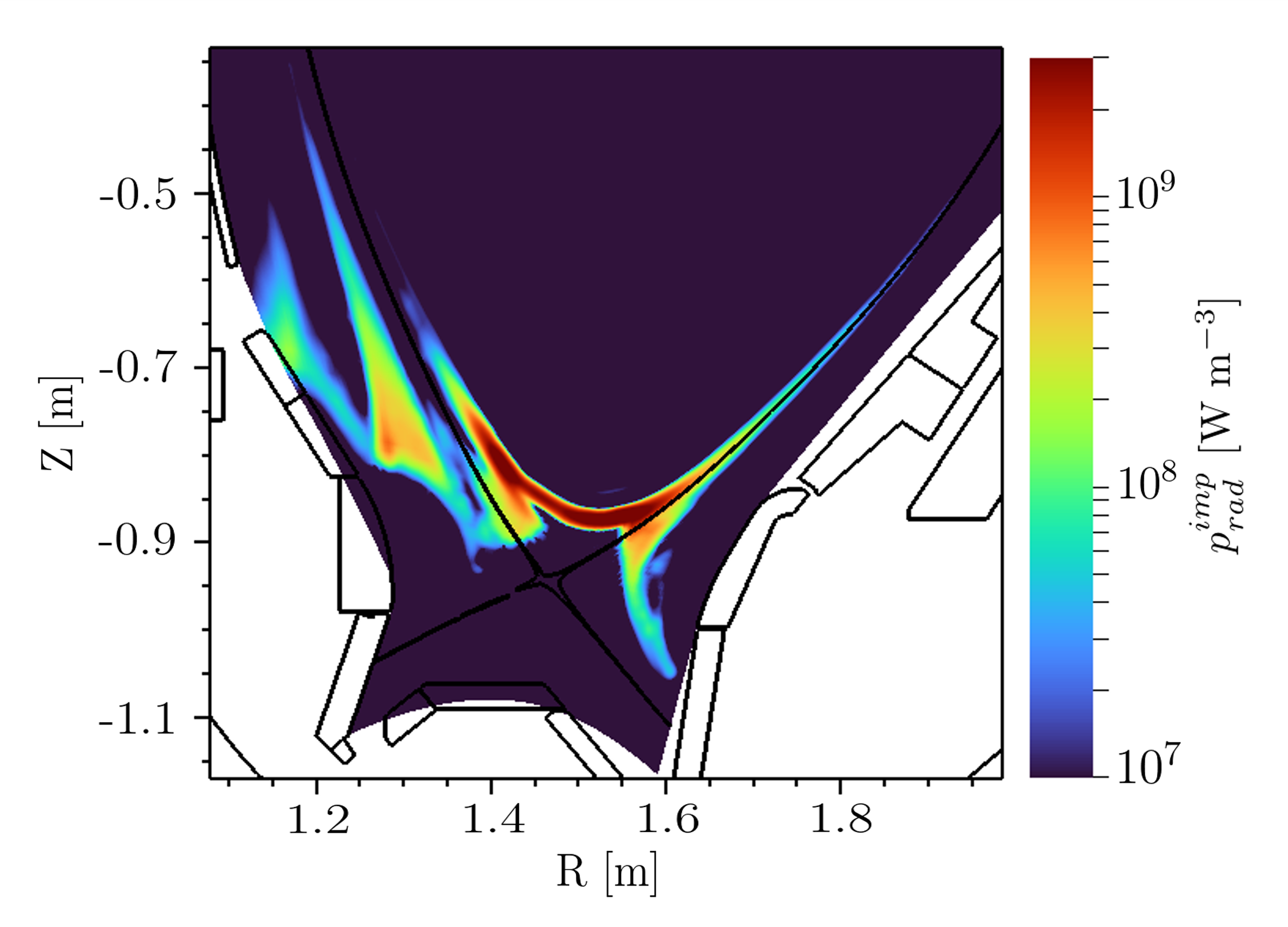}
    \caption{Cross section of the impurity radiation density in the XPR simulation including the impurity collisions.}
    \label{coll-Prad}
\end{figure}

Figure \ref{coll-Prad} shows the cross-section of the impurity radiation distribution of the XPR in the collisional case. Since this collisional term serves as friction against the impurity particles transporting through the background plasma, the impurities are more compressed, leading to a more poloidally localised radiating mantle above the X-point. This shows more similarity with experimental reconstruction~\cite{XPR_Bernert_2017_NME} and SOLPS-ITER work~\cite{SOLPS_Pan_2023_NF}.

However, it is important to note that the inclusion of impurity collisions made the simulation more prone to numerical issues, so the simulation setup (eg. transport coefficients, rates of gas puff) had to be adjusted from the reference case. Therefore, the collisional and the reference case are not a direct comparison between the cases with and without collisions, but they are simply a demonstration of how the collisions generally affect the impurity distribution and consequently the shape of the XPR.

\section{Conclusion and outlook}\label{conclusion}

In this article, the formation and the dynamic properties of the X-point radiator were investigated, which constitutes a possible solution for the challenge of practical heat and particle exhaust in a fusion reactor. Using a hybrid fluid-kinetic model of the JOREK code, axisymmetric (2D) simulations were conducted in the AUG geometry, with the simulation grid extended to the first wall. The fluid-kinetic model features the coupling between the base reduced MHD model of JOREK and the full-f kinetic model. The MHD model describes the background main species (deuterium) plasma, and the kinetic model describes the deuterium neutral particles and impurities (nitrogen). In the presented simulations, it was shown that the model is able to simulate the XPR formation and its time evolution in different experimentally relevant regimes. Before the XPR formation, the processes of the HFSHD formation and the detachment of the divertor targets were observed, in accordance with the experimental findings~\cite{Detachment_Cavedon_2022_NF, Detachment_Potzel_2014_NF}.

After the XPR formation, the XPR was successfully kept quasi-stationary with specific fuelling and impurity seeding rates. Then using the quasi-stationary XPR solution as a reference, the simulation was continued with different seeding rates to investigate the dynamic response of the XPR. Two particularly interesting scenarios were analysed in detail. One is the MARFE case, where an increased seeding rate leads to an unstable XPR. The other is the retreating case, in which the impurity seeding is turned off, leading to a successive downward motion and disappearance of the XPR.

In the reference case, the vertical profile of the XPR was analysed. The peak of the impurity radiation locates at around 6.8\;cm ($\rho_{pol}$\;=\;0.995). Below this radiating mantle is the XPR core ($T_e$\;=\;1--2\;eV), where the neutral density and the impurity density are both around one order of magnitude smaller than the local electron density of about 3\;$\times$\;10$^{20}$\;m$^{-3}$. The impurity concentration in the XPR core is around 10--15\;\%, which is similar to the SOLPS-ITER simulations in TCV~\cite{SOLPS_Sun_2025_arXiv}. Regarding the sink and the source of the deuterium neutrals, the ionisation front is located slightly above the radiation peak, whilst volumetric recombination mainly takes place below it. Furthermore, the simulation showed an electric potential hill above the XPR, and the resulting $E$$\times$$B$ transport leads to a stronger poloidal extension of the dense plasma volume towards the inner midplane. A similar $E$$\times$$B$ vortex was also present in the SOLPS-ITER simulations in Ref.~\cite{SOLPS_Pan_2023_NF} and in the GRILLIX simulations in Ref.~\cite{GRILLIX_Eder_2025_NF}.

In the MARFE case, the upward development of the XPR and the MARFE formation were analysed. As the XPR developed upwards, the following process was observed: the impurity density increases in the radiating mantle $ \mapsto $ the radiation peak moves upwards and cools down the ionisation front further $ \mapsto $ neutrals can penetrate further and the ionisation front moves upwards $ \mapsto $ a new XPR solution with a larger XPR height forms. This is consistent with the assumptions in Ref.~\cite{XPR_model_Stroth_2025_PPCF}. Regarding the MARFE formation; as the radiation peaks more strongly and the radiating mantle overlaps with the ionisation front, the XPR eventually becomes MARFE-like. In the MARFE core, the enhanced recombination reduces the electron density, and the electron temperature also drops to well below 1\;eV.

In the retreating case, the downward development and the loss of the XPR were analysed. As the XPR developed downwards, the following process was observed: the impurity density decreases in the radiating mantle $ \mapsto $ the power dissipation via impurity radiation is weakened $ \mapsto $ the ionisation rate of deuterium neutrals is increased in the radiating mantle as the temperature rises $ \mapsto $ the decrease of the neutral density leads to a new XPR solution with a reduced XPR height. Regarding the loss of the XPR; as the neutral content in the XPR core is depleted through ionisation, the XPR is lost despite no further decrease in the impurity density. The radiating mantle then moves into the downstream HFS SOL, where the impurities and neutrals in the SOL are compressed. Additionally, the electric potential hill is lost along with the XPR.

Finally, it was demonstrated that a poloidally localised XPR could be simulated with this fluid-kinetic model in JOREK. This was achieved via the inclusion of the impurity collisions with the background plasma, which served as friction between the impurities and the background plasma.

Future work that can be built upon the results shown in this manuscript includes:

\begin{itemize}
    \item Improving the setup for quantitative comparisons to the experimental data. (This includes an improved setup of particle/ heat transport coefficients and sources in order to better match the outer midplane profiles with experimental measurement.)
    \item Studying the dynamical response of the outer midplane plasma profiles during the vertical movement of the XPR~\cite{XPR_model_Stroth_2025_PPCF}.
    \item Analysing the power and particle balance in the XPR and comparing the parallel and perpendicular transport.
    \item Studying the pressure balance in the evolution from XPR to MARFE, including all contributions from the background plasma, kinetic neutrals, kinetic impurities and the plasma flows. This will clarify whether the simulations develop a pressure hole in the MARFE core, as suggested in Refs.~\cite{XPR_model_Stroth_2022_NF, MARFE_model_Simakov_2000_PoP}.
    \item Testing whether a different quasi-stationary XPR solution can be maintained by increasing the fuelling rate when the seeding rate is reduced.
    \item Transitioning to 3D simulations, for studying the mutual interaction between MHD instabilities and the atomic physics phenomena. This includes investigating the XPR dynamics across an ELM cycle, the suppression of ELMs by the XPR, and the MHD activity arising in the MARFE scenario. The effect of MHD activities on impurity transport is also an important subject.
    \item Including pumping (for better description of the boundary condition), molecular reactions (for better modelling of the plasma at $T_e$\;$<$\;5\;eV, and potentially easier access to detachment), interactions between the deuterium neutrals and the impurities, and self-collisions for either particle species (for more accurate transport and neutral pressure~\cite{SOLPS_Zito_2025_NF}) in the model.
\end{itemize}

\section*{Acknowledgements}\label{acknowledgement}

The authors gratefully acknowledge Ulrich Stroth and Matthias Bernert for the discussions on the experimental observations of the XPR in AUG. This work has been carried out within the framework of the EUROfusion Consortium, funded by the European Union via the Euratom Research and Training Programme(Grant Agreement No. 101052200—EUROfusion). Views and opinions expressed are however those of the author(s) only and do not necessarily reflect those of the European Union or the European Commission. Neither the European Union nor the European Commission can be held responsible for them.



{\fontsize{9pt}{10pt}\selectfont
\bibliography{0_literature}
}

\end{document}